\documentclass[twocolumn,showpacs,preprintnumbers,amsmath,amssymb]{revtex4}

\usepackage{graphicx}
\usepackage{dcolumn}
\usepackage{bm}

\begin{document}

\preprint{APS/123-QED}

\title{Rank-ordered Multifractal Spectrum for Intermittent Fluctuations}

\author{Tom Chang}
\affiliation{%
Kavli Institute for Astrophysics and Space Research, 
Massachusetts Institute of Technology, Cambridge, MA 02139, USA
}%

\author{Cheng-chin Wu}
\affiliation{
Institute of Geophysics and Planetary Physics, University of California, Los Angeles, CA 90095, USA 	
}%

\date{\today}

\begin{abstract}
We describe a new method that is both physically explicable and quantitatively accurate in describing the multifractal characteristics of intermittent events based on groupings of rank-ordered fluctuations. The generic nature of such rank-ordered spectrum leads it to a natural connection with the concept of one-parameter scaling for monofractals. We demonstrate this technique using results obtained from a 2D MHD simulation. The calculated spectrum suggests a crossover from the near Gaussian characteristics of small amplitude fluctuations to the extreme intermittent state of large rare events.
\end{abstract}

\maketitle

Intermittent fluctuations are popularly analyzed using the structure function and/or partition function methods.  These methods investigate the multifractal characteristics of intermittency based on the statistics of the full set of fluctuations.  Since most of the observed or simulated intermittent fluctuations are dominated by fluctuations with small amplitudes, the subdominant fractal characteristics of the minority fluctuations -- generally of larger amplitudes
--  are easily masked by those characterized by the dominant population.  It therefore appears prudent to search for a procedure that explores the singular nature of the subdominant fluctuations by first appropriately isolating out the minority populations and then perform statistical investigations for each of the isolated populations.  Using an example, we demonstrate how this idea may be achieved with a rank-order method that subdivides the fluctuations into groupings based on the range of the scaled-sizes of the fluctuations.

		To be more specific, we consider a generic fluctuating temporal event $X(t)$ and form from which a scale dependent difference series $\delta X(t,\tau)=X(t+\tau)-X(t)$  for a time lag $\tau$.  We now consider the probability distribution functions (PDFs) $P(\delta X,\tau)$  of $\delta X(t,\tau)$  for different time lag values $\tau$.  If the phenomenon represented by the fluctuating event $X(t)$ is monofractal – 
i.e., self-similar, the PDFs would scale (collapse) onto one scaling function $P_s$  as follows:
\begin{equation}
P(\delta X,\tau)\tau^s=P_s(\delta X/\tau^s) 
\end{equation}                                                                                                  
where $s$  is the scaling exponent.  Such one-parameter scaling has been suggested for the stock market indices [1], magnetic fluctuations [2, 3] and fluctuating events of other natural or experimental systems [4].  
If the PDFs are Gaussian distributions for all time lags similar to those characterizing self-similar random diffusion, the scaling exponent s is equal to 0.5.  For other monofractal distributions, the scaling exponent may
take on any real value.

In practical applications, expression (1) is sometimes approximately satisfied for the full range of the scaling variable $Y\equiv\delta X/\tau^s$ and sometimes only for a portion of the range of $Y$.  If the scaling exponent $s$  is obtained by estimating the values of the PDFs at $\delta X=0$, the PDFs would generally scale at least for a range of $Y$ close to the origin. When scaling of the PDFs based on (1) is not fully satisfied or only approximately satisfied, the fluctuating phenomenon represented by $X(t)$ is multifractal.  One conventional method of evaluating the degree of multifractal (intermittent) nature of $X(t)$ is to study the scaling behavior of the moments of the PDFs (conventionally called the structure functions).
\begin{equation}
S_m(\tau)=\left<\left|\delta X(\tau)\right|^m\right>=\int_0^{\delta X_{\rm max}}
\left|\delta X(\tau)\right|^m P(\delta X,\tau) {\rm d}\delta X
\end{equation}                                                           
where $<...>$ represents the ensemble average and $\delta X_{\rm max}$  is the largest value of $\delta X$ obtainable from the time series $X(t)$ for the time lag $\tau$. The choice of taking the ensemble average of the absolute values of the coarse-grained differences instead of the values of the raw differences is for the purpose of better statistical convergence [5, 6].  One then proceeds to search for the scaling behavior $S_m(\tau)\sim\tau^{\zeta_m}$.  If such scaling is verified for a monofractal fluctuating event $X(t)$,  the structure function exponents would vary linearly with the moment order as $\zeta_m=\zeta_1 m$  where $\zeta_1=s$.  If the structure function exponents deviate from the above linear relationship, the fluctuating event is multifractal.  There are several disadvantages of this approach.  Firstly, the statistical analysis as prescribed above incorporates the full set of fluctuations represented by $X(t)$.  As with most observed PDFs, the statistics are generally dominated by those fluctuations with small amplitudes.  Thus, the fractal (multifractal) nature of the subdominant (larger amplitude) fluctuations is usually masked by the fractal nature of the dominant (smaller amplitude) fluctuations.  Secondly, although deviations of the structure function exponents from the linear form would indicate that the fluctuating event $X(t)$  is multifractal, the physical interpretation of the multifractal nature is not easily deciphered by merely examining the curvature of the deviation from linearity.  Thirdly, the structure functions are usually ill-defined for negative values of m.  We therefore search for a procedure that would remedy the above defects as shown below.

		From the above argument, it appears prudent to perform statistical analyses individually for subsets of the fluctuations that characterize the various fractal behaviors within the full multifractal set.  We recognize that such grouping of fluctuations must depend somehow on the sizes of the fluctuations.  However, we also realize that groupings cannot depend merely on the raw values of the sizes of the fluctuations because the ranges will be different for different scales (time lags $\tau$).  Thus, we proceed to rank-order the sizes of the fluctuations based on the ranges of the scaled variable $Y$, defined above.  For each chosen range of $\Delta Y$  we shall assume that the fluctuations of all time lags will exhibit monofractal behavior and be characterized by a scaling exponent $s$.  The question is then how can this procedure be accomplished.  We continue by constructing the structure functions for the chosen grouping of the fluctuations by performing moment -- structure function -- calculations as prescribed by (2) with the limits of integral replaced by the end points of the range of the chosen $\Delta Y$ for each time lag $\tau$.  We then search for the scaling property of the structure functions that varies as $sm$.   If such scaling property exists, then we have found one region of the multifractal spectrum of the fluctuations such that the PDFs in the range of $\Delta Y$ collapses onto one scaled PDF.  Continuing this procedure for all ranges of $\Delta Y$ produces the rank-ordered multifractal spectrum $s(Y)$ that we are looking for.  The determined values of $s$  for each grouping should be un-affected by the statistics of other subsets of fluctuations that are not within the chosen range $\Delta Y$ and therefore should be quantitatively quite accurate.  The physical meaning of this spectrum is that the PDFs for all time lags collapse onto one master multifractal scaled PDF.  The spectrum is implicit since $Y$ is a defined as a function of $s$.

\begin{figure}
\resizebox{3.3in}{!} {
\includegraphics{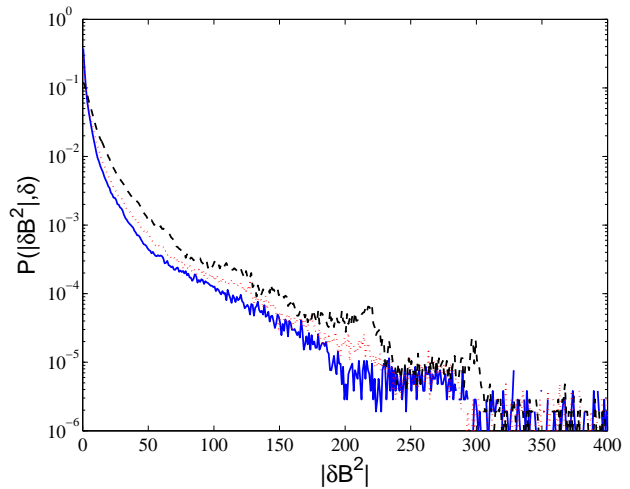} }
\caption{
Raw PDFs as functions of  $|\delta B^2|$ for $\delta = 8$  (solid curve in blue), 16 (dotted curve in red), and 64 (dash curve in black).  $|\delta B^2|$ is in units of bin size chosen as $|\delta B^2|$(max)/800.  There are 800 bins and only first 400 are shown.}
\end{figure}

\begin{figure}
\resizebox{3.3in}{!} {
\includegraphics{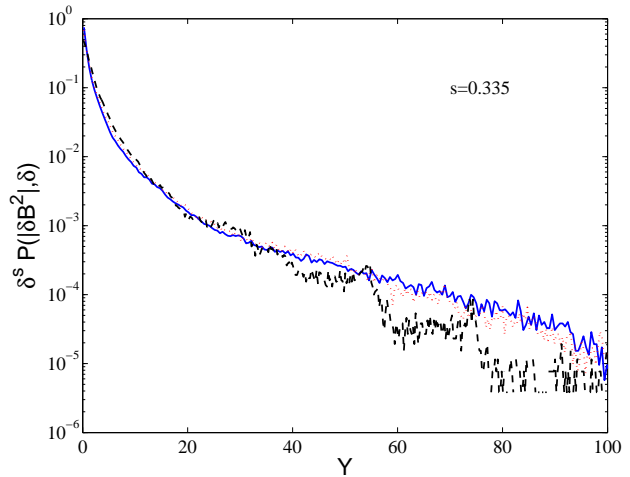} }
\caption{Scaled PDFs with $s=0.335$. Same line styles and colors as in Fig. 1. $Y=|\delta B^2|/\delta^s$; $|\delta B^2|$ in units of bin size and $\delta$ in grid spacing.}
\end{figure}

\begin{figure}
\resizebox{3.1in}{!} {
\includegraphics{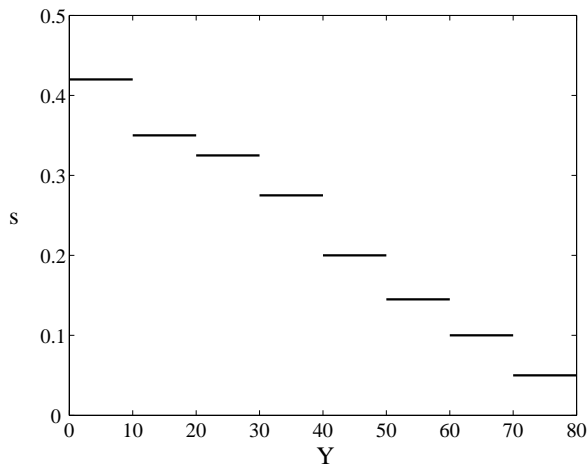} }
\caption{Rank-ordered multifractal spectrum, $s(Y)$.  $Y=|\delta B^2|/\delta^s$;  $|\delta B^2|$ in units of bin size and $\delta$ in grid spacing.  The spectrum is calculated for 8 contiguous ranges of $\Delta Y$.}
\end{figure}

		We demonstrate the above outlined procedure via an example.  The example is based on the results of a large-scale 2D magnetohydrodynamic (MHD) simulation.  In the simulation, ideal compressible MHD equations expressed in conservative forms are solved numerically with 1024 x 1024 grid points in a doubly periodic $(x,z)$ domain of length $2\pi$  in both directions using the WENO code [7] so that the total mass, energy, magnetic fluxes and momenta are conserved.  The initial condition consists of random magnetic field and velocity with a constant total pressure for a high beta plasma.  After sufficient elapsed time, the system evolves into a set of randomly interacting multiscale coherent structures exhibiting classical aspects of intermittent fluctuations.  More detailed description of the simulation was reported in one of our previous publications [3].    Spatial values (approximately one million data points) of the square of the strength of the magnetic field $B^2$ are collected over the entire $(x,z)$ - plane at a given time and PDFs $P(|\delta B^2|, \delta)$  are constructed for the absolute values of the spatial fluctuations  $|\delta B^2|$
at scale intervals $\delta$=8, 16, and 64  grid points, Fig. 1.  Thus, in lieu of temporal fluctuations, the example considers spatial fluctuations.  The PDFs are non-Gaussian and become more and more heavy-detailed at smaller and smaller scales.

		An attempt to collapse the unscaled PDFs according to the monofractal scaling formula that is analogous to (1) indicate approximate scaling with an estimated scaling exponent s=0.335, Fig. 2.  Structure function calculations based on the full set of simulated fluctuations showed a nonlinear relation between the exponents and the moment order [8].  Because for this example the PDFs exhibited approximate monofractal scaling, such structure function calculations -- though indicating multifractality -- is strongly masked by the population of the smaller fluctuations.  Thus, this example represents an ideal candidate to test the utility of the new method described in this paper.

		We now proceed to construct the rank-ordered multifractal spectrum based on the afore-mentioned procedure.  Thus we sort the fluctuations into ranges of $\Delta Y$  between ($Y_1$, $Y_2$) with $Y=|\delta B^2|/\delta^s$  and evaluate the rank-ordered structure functions  within each range:
\begin{equation}
S_m(|\delta B^2|,\delta)=\int_{a_1}^{a_2}|\delta B^2|^m P(|\delta B^2|,\delta) {\rm d}(|\delta B^2|)
\end{equation}                
where $a_1=Y_1\delta^s$ and $a_2=Y_2\delta^s$.  Expression (3) is a nonlinear function of $s$ for each moment order $m$.  We now search for the value(s) of $s$  such that $S_m\sim \delta^{sm}$ within each range of the fluctuations so that the rank-ordered fluctuations would exhibit monofractal behavior.  Interestingly there exists one and only one value of $s$ in each range of $\Delta Y$, that satisfies the above constraint, indicating the appropriateness of the ansatz.  Unlike the structure functions defined for the full range of fluctuations, the range-limited structure functions based on (3) exists also for negative real values of $m$.  Figure 3 displays the calculated rank-ordered spectrum  $s(Y)$ based on eight contiguous ranges of $\Delta Y$.  It is noted that the spectrum has values of $s$ ranging between 0.5 and 0.0.  The spectrum can be refined by choosing more range intervals with smaller range sizes of $\Delta Y$, although in practice this procedure is limited by the availability of simulated data points.  
At $Y=0$, the scaling exponent appears to approach the self-similar Gaussian value of 0.5. As the value of the scaled fluctuation size $Y$ increases the scaling exponent decreases accordingly indicating the fluctuations are becoming more and more intermittent.  At the extremely intermittent state, the value of the scaling exponent  would asymptotically approach the value of zero.  This would occur at the limit of largest and rarest scaled fluctuations.

		For each range of $\Delta Y$, the PDFs would collapse according to its correspondingly calculated exponent value $s$.  For example, for the range of  $Y$ between (40, 50), the PDFs should collapse for the calculated value $s \approx 0.2$.  This is essentially verified as shown in Fig. 4. Figure 5 shows the results of the rank-ordered structure functions for the same range of $\Delta Y$. It also indicates scaling exponent $s =0.2$.

\begin{figure}
\resizebox{3.3in}{!} {
\includegraphics{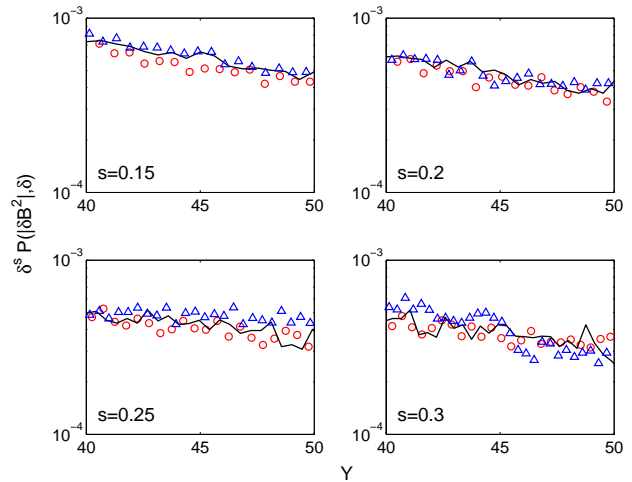} }
\caption{Scaled PDFs for $Y$ in (40, 50) for $s = 0.15$, 0.2, 0.25, and 0.3.  $\delta$ =16 (red open circle), 24(black), 32 (blue open triangle).  Note the triangles are higher than the open circles for both $s$= 0.15 and $s$=0.25.  When $s$=0.2, they are about the same. }
\end{figure}

\begin{figure}
\resizebox{3.3in}{!} {
\includegraphics{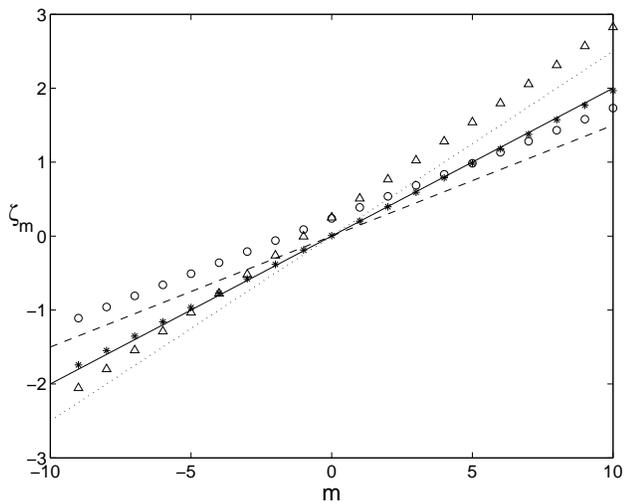} }
\caption{Results of rank-ordered structure functions for $Y$ in (40, 50) for $s = 0.15$ (circle), 0.2 (star), and 0.25 (triangle). $\zeta_m$ is defined by $S_m(|\delta B^2|,\delta) \sim \delta^{\zeta_m}$ and is obtained by a linear fit of log($S_m(|\delta B^2|,\delta)$) versus log($\delta$) for $\delta$ from 16 to 32. The dash line, solid line and dotted line show linear functions $sm$ for s=0.15, 0.2, and 0.25, respectively. For $s$=0.2,  $\zeta_m \approx sm$.}
\end{figure}

		Such an implicit multifractal spectrum has several advantages over the results obtainable using the conventional structure function and/or partition function calculations.  Firstly, the utility of the spectrum is to fully collapse the unscaled PDFs.  Secondly, the physical interpretation is clear.  It indicates how intermittent (in terms of the value of $s$) are the scaled fluctuations once the value of $Y$ is given.   Thirdly, the determination of the values of the fractal nature of the grouped fluctuations is not affected by the statistics of other fluctuations that do not exhibit the same fractal characteristics.  Fourthly, it provides a natural connection between the one-parameter scaling idea (1) and the multifractal behavior of intermittency.  	

		To summarize, we have introduced a new rank-ordered procedure based on the sizes of the scaled fluctuations to view the multifractal nature of intermittent fluctuations.  The suggested implicit multifractal spectrum analysis provides a physically meaningful description of intermittency and is quantitatively accurate because of the cleanliness of the procedure of statistical sampling.  The method can easily be generalized to situations of higher dimensions as well as correlation and response functions of several independent variables involving intermittency of spatiotemporal fluctuations.

This research is partially supported by the NSF and AFOSR.


\end{document}